\documentclass[11pt,a4paper]{article}
\pdfoutput=1

\usepackage{jcappub}
\usepackage{amsmath}
\usepackage{amssymb}

\def\approxprop{
  \def\p{
    \setbox0=\vbox{\hbox{$\propto$}}
    \ht0=0.6ex \box0 }
  \def\s{
    \vbox{\hbox{$\sim$}}
  }
  \mathrel{\raisebox{0.7ex}{
      \mbox{$\underset{\s}{\p}$}
    }}
}

\title{Degeneracy of gravitational waveforms in the context of GW150914}

\author[a]{{James Creswell},}
\author[a,b]{{Hao Liu},}
\author[c]{{Andrew D. Jackson},}
\author[a]{{Sebastian {von Hausegger}},}
\author[a]{and {Pavel Naselsky}}

\affiliation[a]{The Niels Bohr Institute \& Discovery Center, University of Copenhagen, Blegdamsvej 17, DK-2100 Copenhagen, Denmark}
\affiliation[b]{Key Laboratory of Particle and Astrophysics, Institute of High Energy Physics, CAS, 19B YuQuan Road, Beijing, China}
\affiliation[c]{Niels Bohr International Academy, University of Copenhagen, Blegdamsvej 17, DK-2100 Copenhagen, Denmark}
\emailAdd{dgz764@alumni.ku.dk}
\emailAdd{liuhao@nbi.dk}
\emailAdd{jackson@nbi.dk}
\emailAdd{s.vonhausegger@nbi.dk}
\emailAdd{naselsky@nbi.dk}

\abstract{We study the degeneracy of theoretical gravitational waveforms for binary black hole mergers using an aligned-spin effective-one-body model. After appropriate truncation, bandpassing, and matching, we identify regions in the mass--spin parameter space containing waveforms similar to the template proposed for GW150914, with masses $m_1 = 36^{+5}_{-4} M_\odot$ and $m_2 = 29^{+4}_{-4} M_\odot$, using the cross-correlation coefficient as a measure of the similarity between waveforms. Remarkably high cross-correlations are found across broad regions of parameter space. The associated uncertanties exceed these from LIGO's Bayesian analysis considerably. We have shown that waveforms with greatly increased masses, such as $m_1 = 70 M_\odot$ and $m_2 = 35 M_\odot$, and strong anti-aligned spins ($\chi_1=0.95$ and $\chi_2=-0.95$) yield almost the same signal-to-noise ratio in the strain data for GW150914.}


\begin{document}
\maketitle
\flushbottom

\section{Introduction}
\label{sec:1}

Two black holes orbiting each other in a binary system emit gravitational waves, causing the orbit to decay gradually.
As the orbit shrinks, gravitational wave emission increases and peaks when the two black holes merge.
Such black hole mergers release enormous amounts of gravitational radiation, and they are one of the primary targets of gravitational wave observatories.
On 14 September 2015, the Laser Interferometer Gravitational-Wave Observatory (LIGO) made the first detection of gravitational waves.
The signal, called GW150914, was reported to be from a binary black hole with masses $m_1=36^{+5}_{-4}$ and $m_2=29^{+4}_{-4}$ solar masses \cite{Ligo1, Ligo2}.

The radiation observed from a binary black hole (BBH) merger depends on 15 parameters, including the masses and spins of the black holes, the distance to the source, the inclination angle between the system's orbital momentum and the line of sight, the polarisation angle, and the time of coalescence \cite{Cutler}.
Other parameters, such as the eccentricity of the orbit, may also play a role, but are typically neglected.
Accurate determination of these parameters is of interest in astrophysics and cosmology.

To leading order in the post-Newtonian (PN) expansion, the frequency evolution of a BBH gravitational wave is governed by the chirp mass,
\begin{equation}
    \mathcal{M} = \frac{(m_1 m_2)^{3/5}}{(m_1 + m_2)^{1/5}}.
    \label{eq:chirp}
\end{equation}
As we move to higher orders, the individual masses, and then the spins, affect the morphology.
We define the dimensionless spins $\chi_1$ and $\chi_2$, which are the spin magnitudes of the black holes in units of the maximum allowed spin of a Kerr black hole; the $\chi_i$ therefore take values between $-1$ and $1$.
We further define the effective total spin of the binary,
\begin{equation}
    \chi = \frac{m_1 \chi_1 + m_2 \chi_2}{m_1 + m_2}.
    \label{eq:chi}
\end{equation}

The main focus of this paper is the dependence of the waveforms on these parameters, focusing in particular on the features of degeneracy that exist.
The restriction of signals like GW150914 to the 35--350 Hz frequency band truncates their low frequency tails down to 100--200 ms and consequently limits the number of cycles that can be seen during the inspiral phase.
This imposes significant restrictions on the precision with which black hole parameters can be inferred.

In principle, features of the parameter space, including degeneracy, are reflected in the Bayesian approach for parameter estimation used by LIGO \cite{Veitch}.
However, this method is limited by several unfounded assumptions, discussed in detail below.

The effects of waveform degeneracy have been studied before, particularly in the context of constructing a sufficiently large template bank for detection \cite{Cutler,Baird,Ajith2,Cho1,Cho2}.
In this paper, our focus is on the degeneracy of BBH waveforms with masses and spins appropriate for GW150914, using methods that compare the morphology of the waveforms themselves.

The outline of the paper is as follows.
In section \ref{sec:2} we discuss parameter estimation using Bayesian inference and properties of the signal-to-noise ratio, which are two central components of LIGO's data analysis.
In section \ref{sec:3} we define the cross-correlation and matching procedure used for comparing waveforms, and in section \ref{sec:4} we summarize the methods for constructing theoretical waveforms.
In section \ref{sec:5} we exhibit the cross-correlation degeneracy.
In section \ref{sec:6} we extend the comparisons to different frequency domains.
In section \ref{sec:7} we show that the degeneracy in waveform morphology carries over to the signal-to-noise ratio in the Hanford and Livingston strain data for GW150914, and in section \ref{sec:8} we estimate the uncertanties in the GW150914 masses using an assumption-free method.
We conclude in section~\ref{sec:9}.

\section{Characterization of GW150914 and properties of signal-to-noise ratio}
\label{sec:2}
\subsection{Parameter estimation using Bayesian inference}
\label{sec:2.1}

In this section we review the method of determining parameters from BBH waveforms with Bayesian inference, used by LIGO \cite{Ligo2, Veitch}.
Given a theoretical template model that predicts templates $h(\theta)$ in terms of the parameters $\theta$, the posterior probability distribution for $\theta$ is given by Bayes' theorem as the product of the prior probability distribution and a likelihood function
\begin{equation}
    \mathcal{L} \propto \exp \left( -\frac{1}{2} \sum_k \Big \langle h_k(\theta) - s_k \Big | h_k(\theta) - s_k \Big \rangle \right),
    \label{eq:likelihood}
\end{equation}
where the sum is taken over the different detectors in the network indexed by $k$, $s_k$ is the strain data in the $k$-th detector, and $h_k(\theta)$ is the template projected onto the $k$-th detector.
The inner product of two time-domain records is given in terms of their Fourier transforms by
\begin{equation}
    \langle h_1 | h_2 \rangle = 4 \ \mathrm{Re} \int_0^\infty \frac{\tilde{h}_1^*(f) \tilde{h}_2(f)}{S_n(f)} df.
    \label{eq:inner}
\end{equation}
In eq.~(\ref{eq:inner}), the two functions are weighted by the noise power spectral density, $S_n(f)$.
The values of the parameters $\theta$ which maximize the posterior distributions (or, equivalently in the case of flat priors, maximize the likelihood function) are taken to be the most likely parameters characterizing the BBH, and the widths of the posterior distributions characterize the uncertainty in these parameters.

The parameter estimates and statistical uncertainties resulting from such a procedure are based on several assumptions about the data and the noise.
First, the likelihood function in eq.~(\ref{eq:likelihood}) assumes Gaussian noise, uncorrelated between detectors and characterized by a known and stationary power spectral density (PSD).
Real LIGO noise violates all of these assumptions, sometimes severely \cite{Martynov,Edwards,Littenberg1,Littenberg2,Liu,Creswell}.

In practice, the PSD is estimated by averaging segments of off-source data (see section~\ref{sec:2.3}).
However, the sampling error associated with this estimate is neither accounted for in the likelihood function nor reflected in the posterior distributions.
Furthermore, the non-stationarity of the noise increases the error associated with PSD estimation.
The inner product in eq.~(\ref{eq:inner}), written as an integral in frequency space, obscures the position of the data in the time domain. 
Although the presence of $S_n(f)$ in the denominator suppresses contributions from regions of poor detector sensitivity, it does not necessarily eliminate them, and therefore the integral can be contaminated by parts of the record where the signal is dominated by noise.

The contributions from each detector are combined in a coherent way that fails to take advantage of redundancy.
For example, in the two detector case, if the data is not at all informative in one detector, the shape of the likelihood function is determined entirely by the other detector.

Finally, the posterior distributions depend strongly on the prior distributions chosen, especially for spins \cite{Vitale}.
The correct forms of the prior distributions are a matter of debate.\footnote{Needless to mention there are also some technical problems of the likelihood analysis related to inversion of the noise covariance matrix $c_{ij}$. The uncertaintiy of the widely used ``diagonal approximation'', where $c_{ij}^{-1} = S_n^{-1} \delta_{ij}$, needs more detalied investigation.}

In summary, a number of questions remain regarding the validity of the likelihood function in eq.~(\ref{eq:likelihood}).
Bayesian inference is performed in the context of a probabilistic model, and if the model does not describe the data generating process accurately, the results are suspect.
The reliability of the Bayesian method can be tested with injections where the signal parameters are known (e.g. ref.~\cite{Veitch}).
Although the chirp mass is recovered accurately, the individual masses and the spins can disagree with the derived posterior distributions.
Some attempts have been made to fortify the inference procedure against the risky assumptions that enter it \cite{Edwards,Littenberg1,Littenberg2}.
However, these modifications do not provide a truly conservative, assumption-free analysis, and they are not mentioned in the GW150914 parameter estimation paper~\cite{Ligo2}.

Inspired by these difficulties with the Bayesian approach, in this paper we study the dependence on black hole parameters of gravitational waves similar to GW150914, and the corresponding limits on the accuracy with which the GW150914 parameters can be determined, using a simple method that makes no assumptions about noise properties.
We manually restrict our analysis to the frequency domain where the detectors are sensitive and the time domain where the signal is strong.

\subsection{Signal-to-noise ratio}
\label{sec:2.2}

It is worth noting that under the assumption of Gaussian noise and flat priors, the best-fit template according to the maximum of the posterior distribution also maximizes the signal-to-noise ratio \cite{Cutler}.\footnote{However, the two methods give different results for GW150914 \cite{Ligo2,Ligo4}. The maximum SNR template has $m_1=48 M_\odot, m_2=37 M_\odot$, which lie outside the credible intervals according to the likelihood function, $m_1=36^{+5}_{-4} M_\odot, m_2=29^{+4}_{-4} M_\odot$.}
As stated above, the noise is not Gaussian and corresponding changes to the methods are required.
Below we define the signal-to-noise ratio (SNR), and in the following subsections we elaborate on aspects of SNR calculations. 

\begin{figure}[tbp]
    \centerline{
    \includegraphics[width=0.5\textwidth]{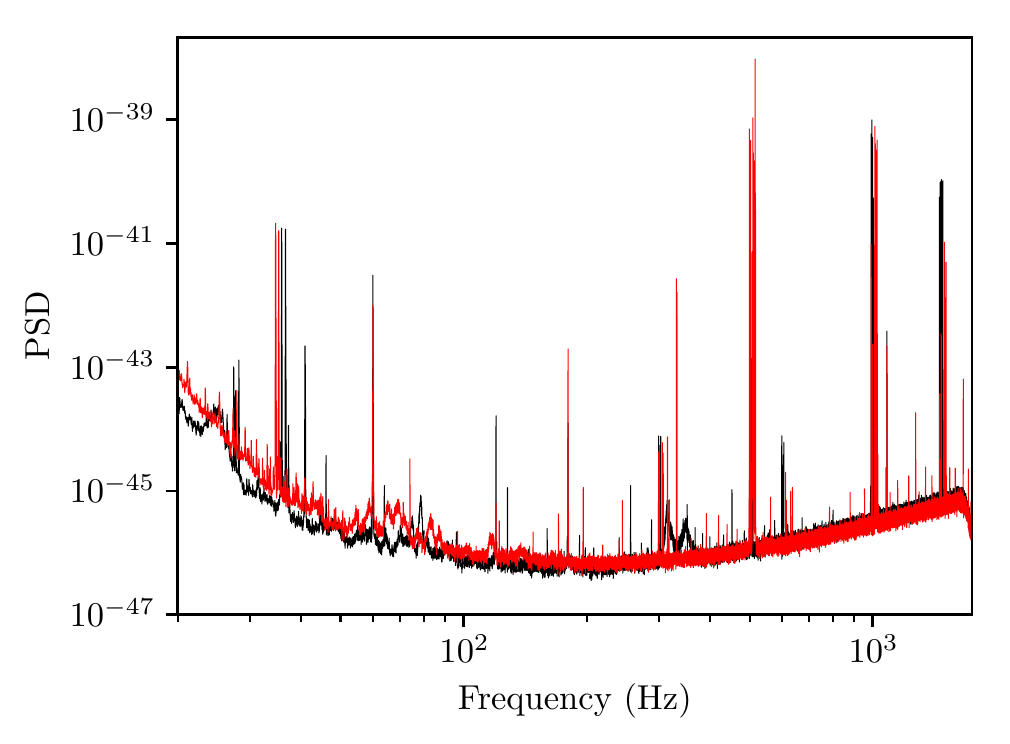}
    \includegraphics[width=0.5\textwidth]{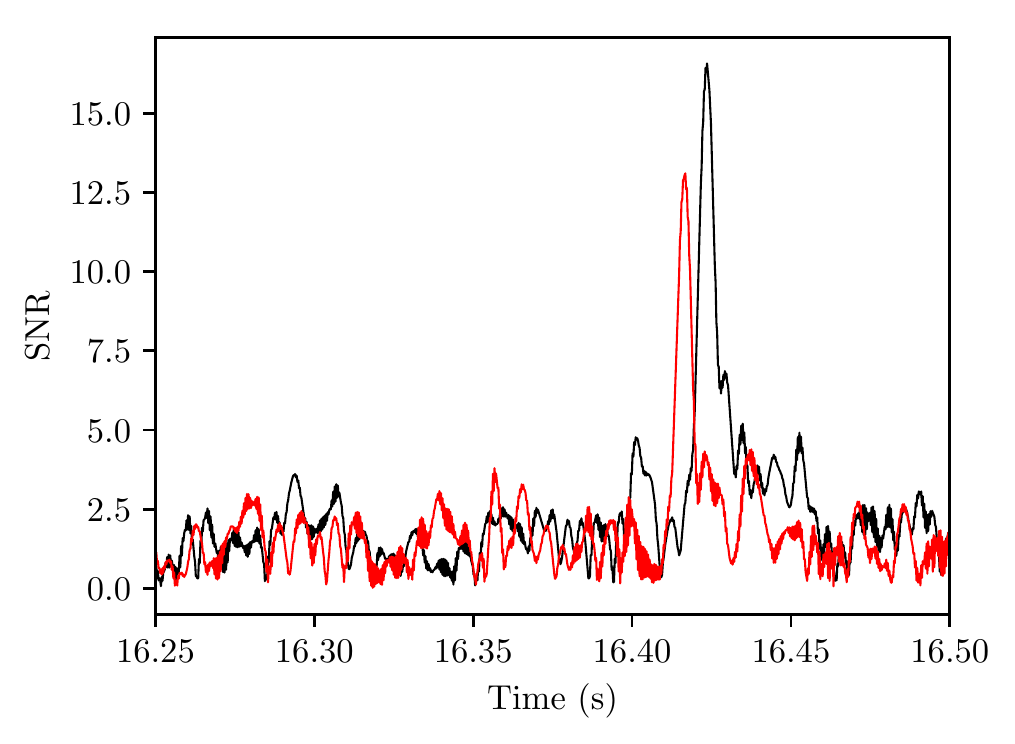}}
    \caption{Characterization of GW150914. Left panel: the PSD $S_n(f)$ in the 2048 seconds surrounding the event. Right panel: the SNR in 32 seconds for the template with $m_1 = 36 M_\odot$, $\chi_1 = 0$, and  $m_2=29 M_\odot$, $\chi_2=0$. Black is Hanford and red is Livingston.}
    \label{fig:1}
\end{figure}

The SNR measures the strength of a proposed signal $h(t)$ in noisy data $s(t)$.
It is defined as
\begin{equation}
    \rho^2(\tau) = \frac{\big | \big \langle s(t) \big | h(t - \tau) \big \rangle \big |^2}{\big \langle h(t) \big | h(t) \big \rangle},
    \label{eq:rho}
\end{equation}
where $\tau$ is a time shift measuring the displacement of the template with respect to the data.
The SNR is often maximised over all possible time shifts, i.e.
\begin{equation}
    \rho = \max_\tau \rho(\tau),
\end{equation}
and the result is called the SNR of a certain template $h(t)$. 
The SNR in each detector as a function of the time shift $\tau$ for the GW150914 data and template is plotted in the right panel of figure~\ref{fig:1}.

Differences in the parameters that affect the frequency evolution of the wave, like the masses and spins, will inevitably cause templates to become out of phase as one considers more cycles during the inspiral.
Therefore, when many cycles of data are present, the SNR is highly sensitive to these parameters.
However, for events like GW150914 where only a few cycles are visible---whether due to larger black hole masses or a weaker signal---the sensitivity to differences in masses is diminished.

\subsection{PSD estimation}
\label{sec:2.3}

We model the strain data in a detector by
\begin{equation}
    s(t)=g(t)+n(t)
    \label{eq:strain}
\end{equation}
where $g(t)$ is a gravitational wave signal and $n(t)$ is the detector noise. 
We are interested in determining the parameters that characterize $g(t)$.
The Bayesian method described in section \ref{sec:2.1} models the noise as a stationary Gaussian process characterized by a PSD, $S_n(f)$.
In practice, however, the actual realization of the noise simultaneous with a gravitational wave event is unknown, and the PSD must be estimated from off-source noise.

LIGO has implemented the following algorithm to estimate the PSD: Divide a certain length, $L$, of the data into small segments of length $l$, and then calculate the PSD for each small segment.
Then, $S_n(f)$ is given by the median over all segments \cite{Veitch,Allen}.
Using $L=2048$~s and $l=32$~s, the result is plotted in the left panel of figure~\ref{fig:1} for the GW150914 event. 
Note that none of the narrow lines (e.g. $60,120,180$ Hz) have been removed from $S_n(f)$.
Since $S_n(f)$ appears in the denominator of eq.~(\ref{eq:inner}), these lines will remove contributions from the corresponding regions in the frequency domain when computing the likelihood function or the SNR.

As seen in figure~\ref{fig:1}, the PSDs in the Hanford and Livingston detectors are different.
Furthermore, during different observational seasons the properties of the PSDs change due to upgrading of the instruments or systematic effects. For instance, for GW150914 the PSD for Hanford
is significantly smaller than Livingston for $f\le25$ Hz. However, for GW170608, the Hanford noise significantly exceeds the noise in Livingston, especially for $f<25$ Hz \cite{Ligo6}. 

\begin{figure}[tbp]
    \centerline{
    \includegraphics[width=0.48\textwidth]{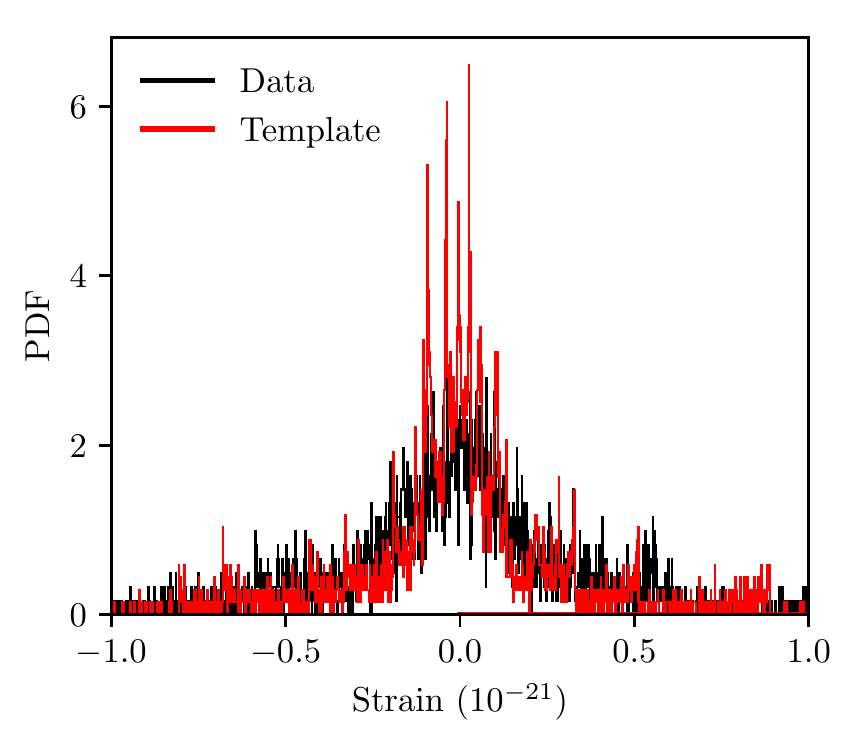}
    \includegraphics[width=0.48\textwidth]{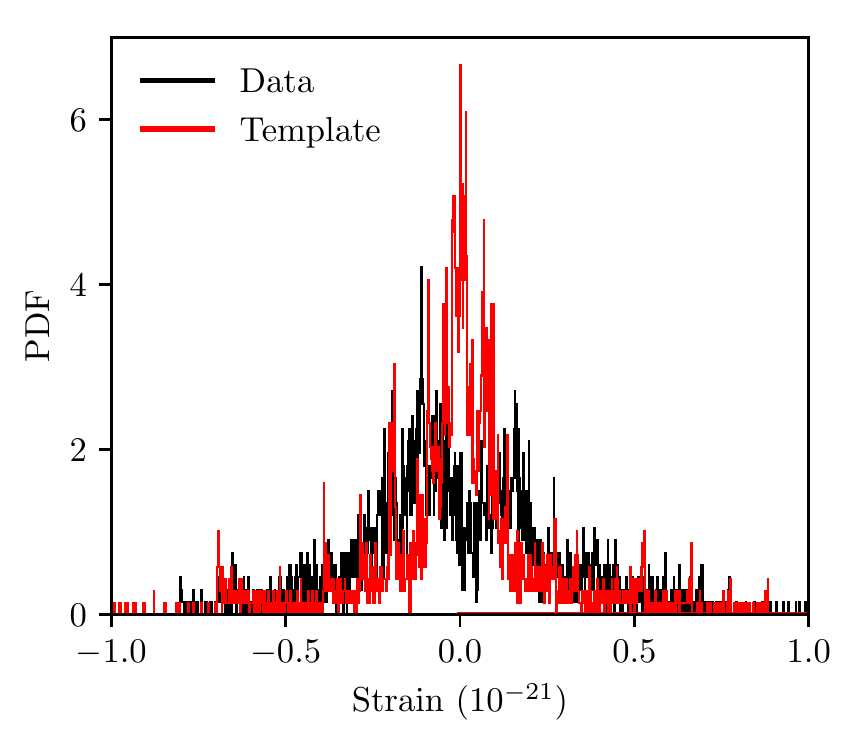}}
    \caption{The normalized number of counts (PDF) for the Hanford (left) and Livingston (right) 200~ms strain data (black) and templates (red) from ref. \cite{LOSC}.}
    \label{fig:2}
\end{figure}

The complications of the noise properties can be illustrated
for the GW150914 event in the publicly available 200 ms cleaned data and templates \cite{LOSC}. In figure \ref{fig:2} we show two histograms for the Hanford and Livingston strains, showing 
the number of counts versus the strain.
We can clearly see that the properties of the noise $n(t)$ in the 200 ms time domain are quite complicated, even if we simply compare the distribution of the Livingston strain data to that of Hanford.

The average PSD, constructed as described above, is not necessarily representative of the power of the noise in the single segment of a gravitational wave event.
The effects of sample variance on the extracted parameters have been investigated in refs. \cite{Aasi} and \cite{Raymond}, and in the context of a full parameter search, rudimentarily parameterized in ref. \cite{Veitch}. 

\subsection{Analytical properties of the SNR}

Moving to the Fourier domain, we define the amplitudes and the phases for the strain and template as
\begin{equation}
    s(\omega)=|s(\omega)|e^{i\Phi_s(\omega)},\quad
    h(\omega)=|h(\omega, \theta)|e^{i\Phi_h(\omega,\theta)}.
\end{equation}
Then, the SNR from eq.~(\ref{eq:rho}) is given in terms of the inner product as
\begin{equation}
    \rho(\tau) \propto \langle s(t) | h(t - \tau) \rangle = 4 \ \mathrm{Re} \int_{0}^{\infty}
    \frac{|h^*(\omega,\theta)||s(\omega)|}{S_n(\omega)}e^{i\Pi(\omega,\tau)} d\omega,
\end{equation}
where
\begin{equation}
    \Pi(\omega,\tau)=\Phi_s(\omega)-\Phi_h(\omega,\theta)+\omega \tau.
    \label{eq:phase}
\end{equation}
We estimate this integral analytically using the stationary phase method.
The phase term 
$\Pi(\omega,\tau)$ in eq.~(\ref{eq:phase}) is stationary when $d\Pi/d\omega=0$ and $d^2\Pi/d\omega^2 > 0$, say at the frequency $\omega_*$.
Expanding  $\Pi(\omega,\tau)$ as a Taylor series around 
$\omega_*$ and neglecting  terms of order higher than $(\omega-\omega_*)^2$
we get:
\begin{equation}
    \rho(\tau) \approxprop 4\frac{|h(\omega_*,\theta)||s(\omega_*)|}{S_n(\omega_*)}\sqrt{\frac{2\pi}{|\Pi''(\omega_*)|}}\cos\left(\Pi(\omega_*,\tau)\pm\pi/4\right)
\label{eq:phase2}
\end{equation}
where
\begin{equation}
    \Pi''(\omega_*)=(\Phi''_s(\omega)-\Phi''_h(\omega,\theta))|_{\omega=\omega_*}.
    \label{eq:phase3}
\end{equation}
From eqs. (\ref{eq:phase2})--(\ref{eq:phase3}) we see that $\rho(\tau)$ has a resonance shape in the vicinity of its maximum\footnote{In the case of multiple maxima, eq. (\ref{eq:phase2}) is valid for each, and $\rho$ is a sum over all maxima.}, where $\Pi''(\omega,\tau)\rightarrow 0$ as $\omega\rightarrow \omega_*$.

In the time domain, where the source $g(t)\gg n(t)$ (see eq. (\ref{eq:strain})), the 
phase difference $\Phi_s(\omega)-\Phi_h(\omega,\theta)$ is given by
\begin{equation}
    \Phi_s(\omega)-\Phi_h(\omega,\theta) \approx \Phi_g(\omega)-\Phi_h(\omega,\theta)+ \frac{|n(\omega)|}{|g(\omega)|}\sin(\psi(\omega)-\Phi_g(\omega))
    \label{eq:phase4}
\end{equation}
where $|n(\omega)|$ and $\psi(\omega)$ are the amplitude and phase
of the noise in the vicinity of the frequency $\omega_*$. 
The second derivative of the phase difference is given mainly by the variation of the noise phase. This is a random variable, dependent on the actual realisation of the noise in the time domain occupied by the true signal.
Note that this phase difference depends not on the PSD, but on the actual realization of the noise during the GW signal, i.e. $n(\omega)$.
Moreover, this phase difference $\Pi(\omega)$ contains all information about the morphology of the signal through the phase of gravitational wave $\Phi_g(\omega)$ and the noise phases $\psi$.
At the same time, the phases of the template $\Phi_h(\omega,\theta)$ can be almost identical for different parameters of the templates $\theta$.
This is why a method independent of a noise model should be considered for the comparison of different templates, complementary to eq.~(\ref{eq:inner}).

\section{Cross-correlation and matched filtering}
\label{sec:3}

Cross-correlation can be used to measure the similarity of two templates.
If $h(t)$ and $g(t)$ are two templates, their Pearson cross-correlation coefficient is \cite{Creswell,Naselsky}:
\begin{equation}
    \mathrm{Corr}(h, g) = \frac{\sum_i (h_i - \bar{h}) (g_i - \bar{g})}{\sqrt{\sum_i (h_i - \bar{h})^2 \sum_i (g_i - \bar{g})^2 }},
    \label{eq:cc}
\end{equation}
where the templates are sampled at equally spaced discrete points labeled by $h_i = h(t_i)$ and $g_i = g(t_i)$.
The sums extend over all points in a given time range of interest, and $\bar{h}$ and $\bar{g}$ are the averages of $h_i$ and $g_i$ within that range.

The cross-correlation measures the similarity in morphology between two templates in a way that is independent of their amplitudes.
Before computing the cross-correlation of two templates, we allow ourselves to ``match'' the templates by introducing a translation in the time domain, a phase in the frequency domain, and an overall scaling. That is, given a waveform $h(t)$, we shall modify its Fourier transform $\tilde{h}(\omega)$ into
\begin{equation}
    \tilde{h}'(\omega) = \alpha \tilde{h}(\omega) e^{i (\Delta + \omega \tau)},
    \label{eq:matching}
\end{equation}
where the parameters $\alpha$, $\Delta$, and $\tau$ are optimized to maximize the cross-correlation of the matched time-domain waveform $h'(t)$ with some specified reference waveform $g(t)$.
Matching waveforms accounts for variations in the source distance, source direction, and arrival time:
Adjusting $\alpha$ corresponds to changing the distance from the source to the detector, and adjusting $\tau$ corresponds to changing the arrival time of the wave.
Neither of these are physical properties of the source itself, so it is sensible to consider waveforms to be equivalent for the purpose of extracting source parameters if they differ only through these parameters.
Adjusting $\Delta$ corresponds to changing the relative orientation of the binary orbit axis with the line of sight and orientation of the detector.
Because the orientation of the binary is unknown, waveforms which only differ through $\Delta$ should also be considered equivalent.

\begin{figure}[tbp]
    \centering
    \includegraphics[width=0.48\textwidth]{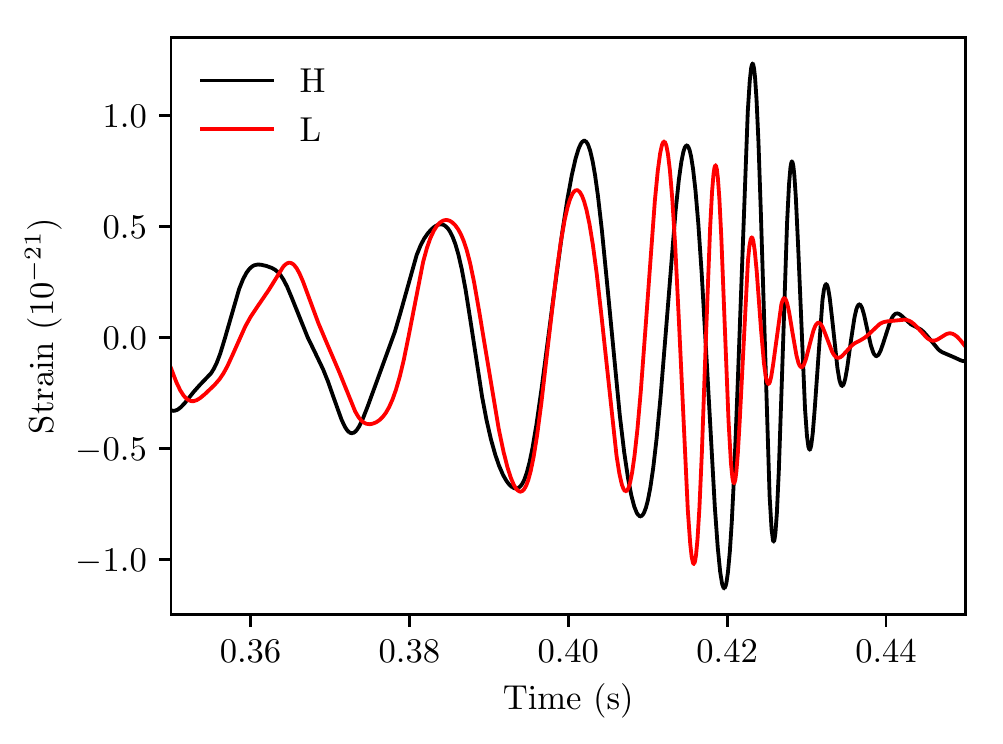}
    \includegraphics[width=0.48\textwidth]{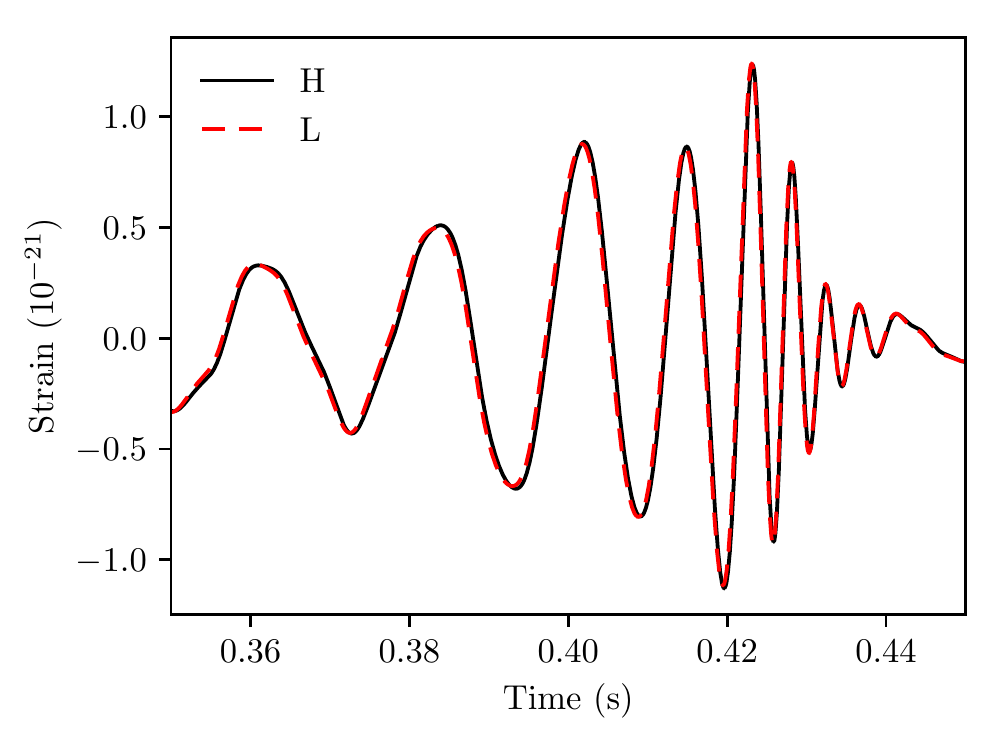}
    \caption{Although the Hanford and Livingston templates appear different, they can be made the same by matching. In the left panel are the original templates, with $\mathrm{Corr}(H, L) = 0.427$. In the right panel Livingston has been matched to Hanford, and then $\mathrm{Corr}(H, L) = 0.999$.} 
    \label{fig:3}
\end{figure}

As an illustration of matching, we consider GW150914 and in particular the published templates \cite{LOSC} in figure 1 of ref.~\cite{Ligo1}.
The Hanford and Livingston templates as they appear are obviously different, and their cross-correlation is $0.427$.
However, by modifying the Fourier transform of the Livingston template as in eq.~(\ref{eq:matching}) with $\alpha = 1.23$, $\Delta = 2.72$, and $\tau = 0.00701 \ \mathrm{s}$, the templates can be made virtually identical, as illustrated in figure \ref{fig:3}.

\section{Waveform models}
\label{sec:4}

An interferometer measures a strain $h(t)$ that is a linear combination of the two gravitational wave polarisations, i.e.
\begin{equation} \label{eq:2.1}
    h(t) = F_+ h_+(t) + F_\times h_\times(t),
\end{equation}
where $F_+$ and $F_\times$ are coefficients depending on the polarisation of the wave and the location of the source relative to the interferometer, and $h_+(t)$ and $h_\times(t)$ are the plus and cross components of the wave.
 
In this paper, we use the \texttt{PyCBC} software package \cite{Canton,Usman,pycbc-software} to generate $h_+(t)$ and $h_\times(t)$ from BBH sources. \texttt{PyCBC} is a Python package with methods for studying and detecting signals from compact binary coalescencses (CBC), based on the LIGO Algorithm Library used by LIGO for the analysis of GW150914 and other gravitational wave events.
Generating full NR waveforms is computationally expensive.
So, for practical purposes it is necessary to use one of several possible approximants.
Here we use the effective-one-body (EOB) approach~\cite{Buonanno}.
In this approach, the motion of two bodies is transformed into the motion of a single effective body in a certain modified Kerr spacetime.
The details of the spin transformation and the background spacetime are parametrized by variables representing unknown higher order PN terms that take values according to calibration with NR simulations.
Specifically, we select the \texttt{SEOBNRv2} model presented in ref. \cite{Taracchini}, which makes the simplifying assumption that the black hole spins are aligned with the orbit axis.
This is equivalent to neglecting precession of the binary.
\texttt{SEOBNRv2} is calibrated to 38 non-precessing NR templates from the SXS collaboration with $1 \leq m_1/m_2 \leq 8$ and $-0.98 \leq \chi_i \leq 0.98$ \cite{Mroue}.
A reduced-order version of \texttt{SEOBNRv2} has been used by LIGO during the O1 observing run to generate templates for BBH searches \cite{Ligo2,Ligo3,Purrer}.

We note that with precession, orbital eccentricity, and higher-order modes neglected, the plus and cross polarisations of a gravitational wave are related in the Fourier domain by $\tilde{h}_+ \propto i \tilde{h}_\times$.
Consequently they are equivalent to each other after matching, and both are equivalent to the strain.
In the following, the ``waveform'' of a gravitational wave is taken to refer to any of these morphologically equivalent objects.

Below we summarize our method for comparing waveforms using cross-correlations and matched filtering, used in section \ref{sec:5} below:
\begin{enumerate}
    \item Generate a reference waveform $h(t)$.
    \item Generate one or more comparison waveforms $g(t)$.
    \item Crop the waveforms to a 100 ms window extending 70 ms before the peak and 30 ms after the peak, or a 200 ms window extending 170 ms before the peak and 30 ms after. These are the primary regions of interest for comparing signals similar to GW150914~\cite{Naselsky}.
    \item Apply a bandpass filter to the waveforms, retaining 35--350 Hz, using a fourth-order Butterworth filter.
    \item For each comparison waveform $g(t)$, optimize the parameters $\Delta$ and $\tau$ to maximize $\mathrm{Corr}(h, g)$.
\end{enumerate}

\section{Comparison of waveforms using cross-correlations}
\label{sec:5}

\subsection[Degeneracy between \texorpdfstring{$m_1$}{m1} and \texorpdfstring{$m_2$}{m2}]{\boldmath Degeneracy between $m_1$ and $m_2$}
\label{sec:5.1}

\begin{figure}[tbp]
    \centering
    \includegraphics[scale=0.8]{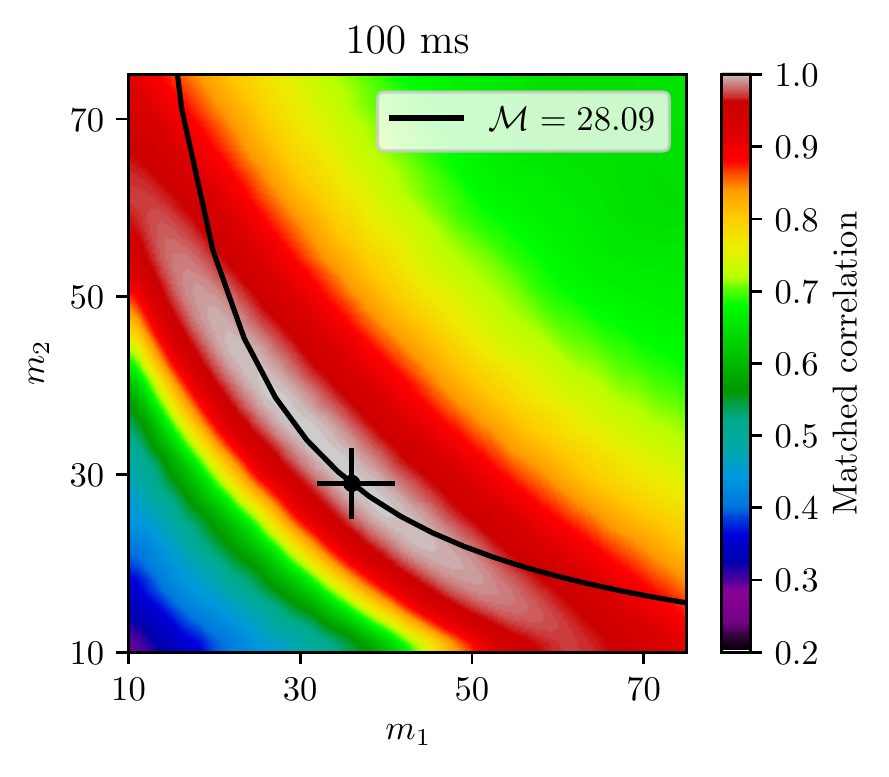}
    \includegraphics[scale=0.8]{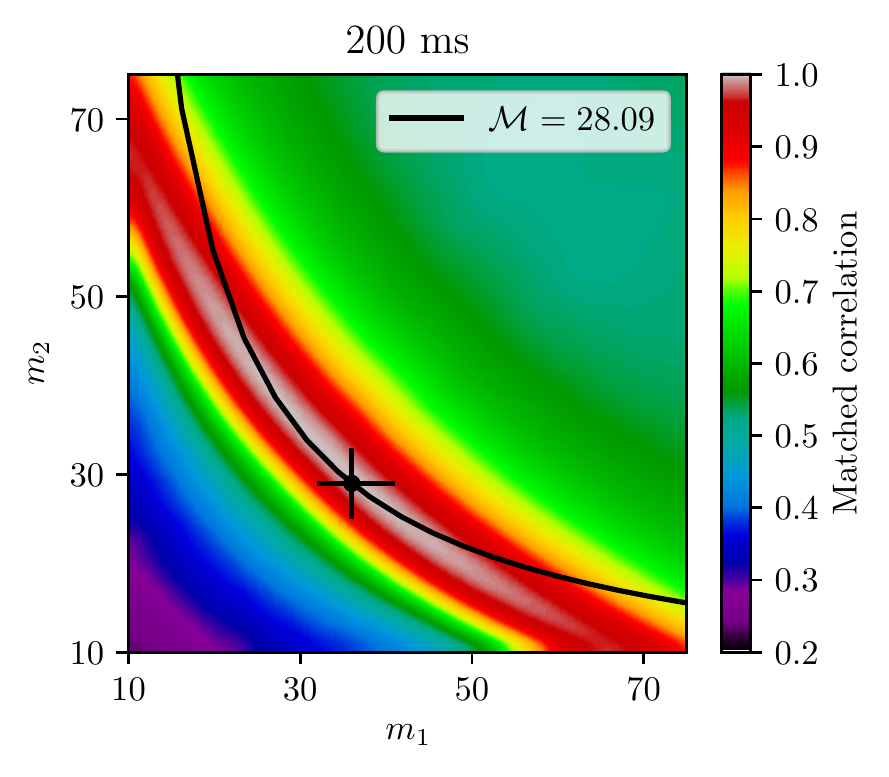}
    \caption{The matched correlation of the GW150914 template with waveforms from the merger of two Schwarzschild black holes with masses $m_1$ and $m_2$, for the 100 ms (left) and 200 ms (right) regions. The templates are bandpassed from $35-350$ Hz. The curve of constant chirp mass $\mathcal{M} = 28.09$ (GW150914) is plotted. The crosses in the middle of each panel indicate the position of $m_1 = 36$, $m_2 = 29$ for GW150914. The size of the crosses correspond to the error bars for determination of the masses $m_1$ and $m_2$ \cite{Ligo1,Ligo2}.}
    \label{fig:4}
\end{figure}

We first set the spins arbitrarily to zero and compare waveforms for various values of $m_1$ and $m_2$.
Since we are interested in waveform degeneracy in the vicinity of GW150914, we take the spinless GW150914 template for our reference waveform, with median parameters $m_1 = 36$, $m_2 = 29$, and $\chi_1 = \chi_2 = 0$.
(Here and below, all masses are given in terms of solar masses.)
We then vary $m_1$ and $m_2$ independently, obtaining all the waveforms for the Schwarzschild case in the mass range $10 < m_1, m_2 < 75$, and calculate the corresponding cross-correlations.
The results are shown in figure \ref{fig:4}.
These plots reveal a remarkably broad range of degeneracy over the entire mass range considered.
The grey region of the map, corresponding to cross-correlations above $0.98$, extends over almost the entire range, including regions far from the cross that indicates the location of the reference template.
This grey region follows the Newtonian degeneracy in eq. (\ref{eq:chirp}) and indicated by the black lines in figure \ref{fig:4}.
The broader zones of near-degeneracy are a consequence of higher order terms, matching, and bandpass filtering.

\subsection[Degeneracy between \texorpdfstring{$\chi_1$}{chi1} and \texorpdfstring{$\chi_2$}{chi2}]{\boldmath Degeneracy between $\chi_1$ and $\chi_2$}
\label{sec:5.2}

\begin{figure}[tbp]
    \centering
    \includegraphics[scale=0.8]{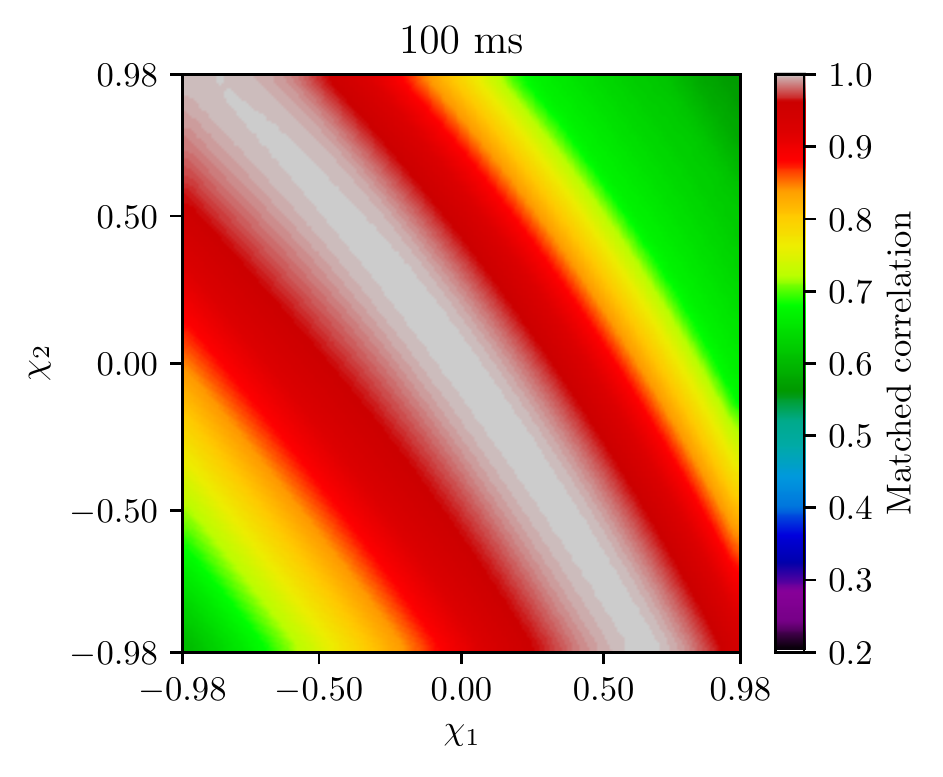}
    \includegraphics[scale=0.8]{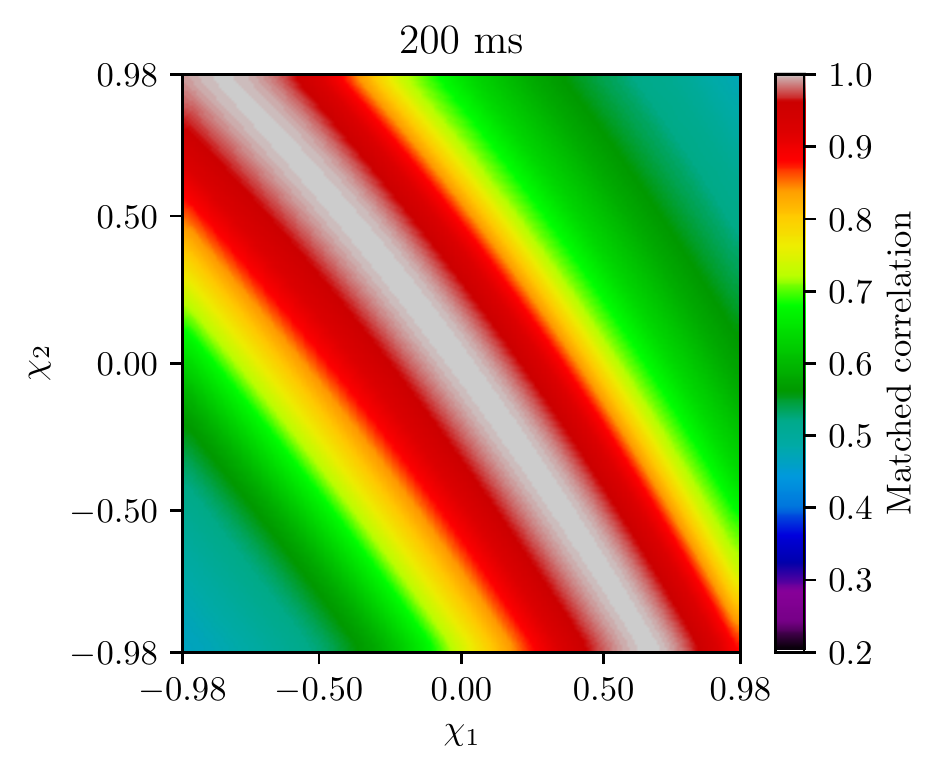}
    \caption{The matched correlation of the GW150914 template with waveforms from the merger of two black holes with masses $m_1 = 36$, $m_2 = 29$ and $z$-axis spins $\chi_1$, $\chi_2$. The left and right panels are for 100 ms and 200 ms respectively.} 
    \label{fig:5}
\end{figure}

Next we consider the comparisons of waveforms with masses fixed at the values of the reference waveform ($m_1 = 36$, $m_2 = 29$) when we vary their spins.
The results are shown in figure~\ref{fig:5}.
As with the masses in the Schwarzschild case above, the spins of the Kerr black holes are not tightly constrained by the morphology.
Note that symmetry with respect to interchanging the two spins is broken as a consequence of the inequality of the two masses.

\subsection[Degeneracy between \texorpdfstring{$m_2$}{m2} and \texorpdfstring{$\chi_2$}{chi2}]{\boldmath Degeneracy between $m_2$ and $\chi_2$}
\label{sec:5.3}

\begin{figure}[tbp]
    \centering
    \includegraphics[scale=0.8]{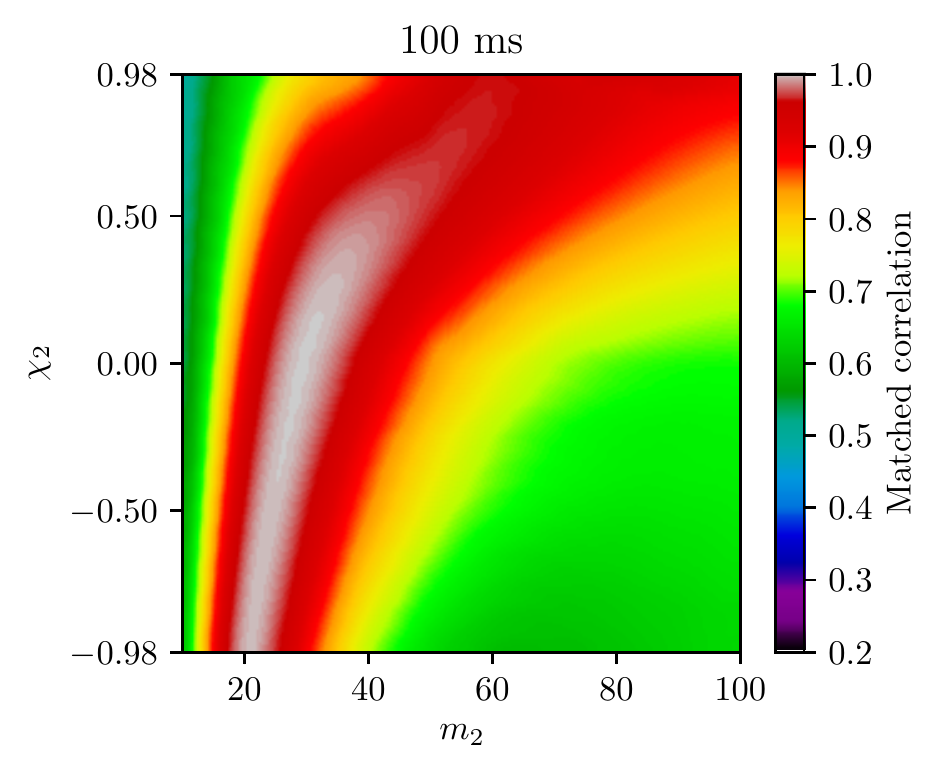}
    \includegraphics[scale=0.8]{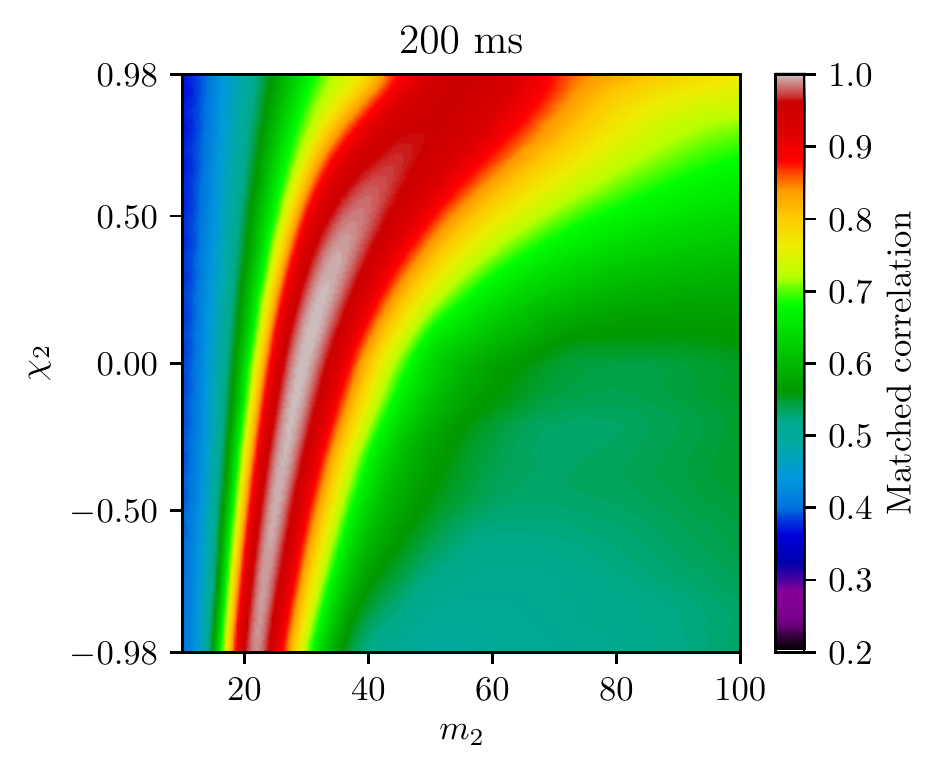}
    \caption{The matched correlation of the GW150914 template with waveforms having $m_1 = 36$, $\chi_1 = 0$, and various values of $m_2$, $\chi_2$. The left and right panels are for 100 ms and 200 ms respectively.}
    \label{fig:6}
\end{figure}

In this model we fix one of the black holes of the comparison waveform with the reference values of $m_1 =36$ and $\chi_1 = 0$, and we vary the mass and spin of the other black hole.
The results are shown in figure \ref{fig:6}.
High cross-correlations are found when the Kerr black hole has a mass less than $m_1$ and a negative spin.
The degeneracy is asymmetric with respect to the zones with $m_2>m_1$.

\subsection[Degeneracy between \texorpdfstring{$M$}{M} and \texorpdfstring{$\chi$}{chi}]{\boldmath Degeneracy between \texorpdfstring{$M$}{M} and \texorpdfstring{$\chi$}{chi}}
\label{sec:5.4}

\begin{figure}[tbp]
    \centering
    \includegraphics[scale=0.8]{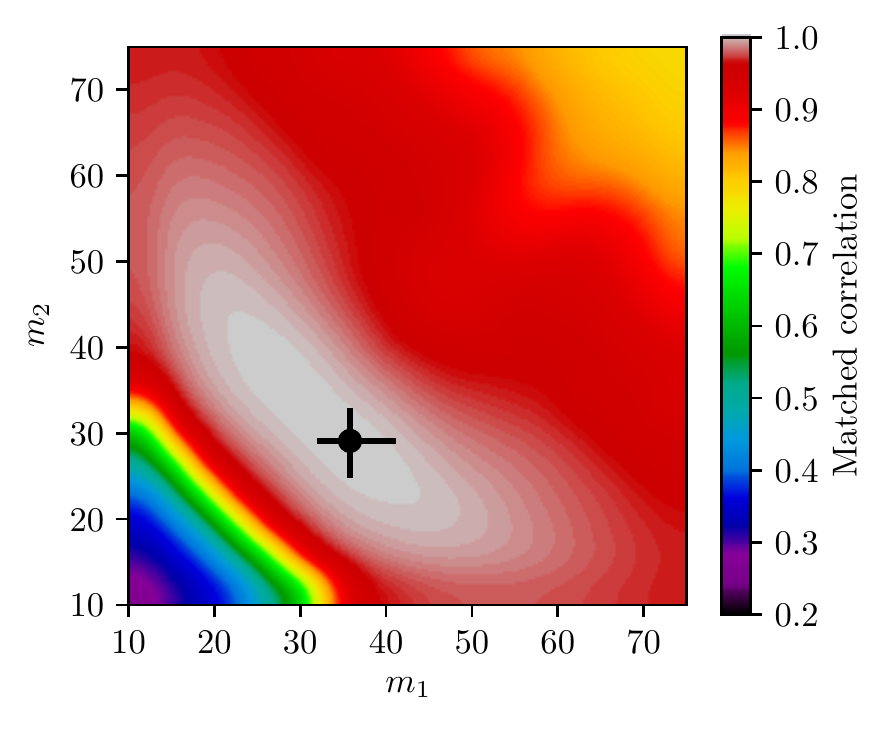}
    \includegraphics[scale=0.8]{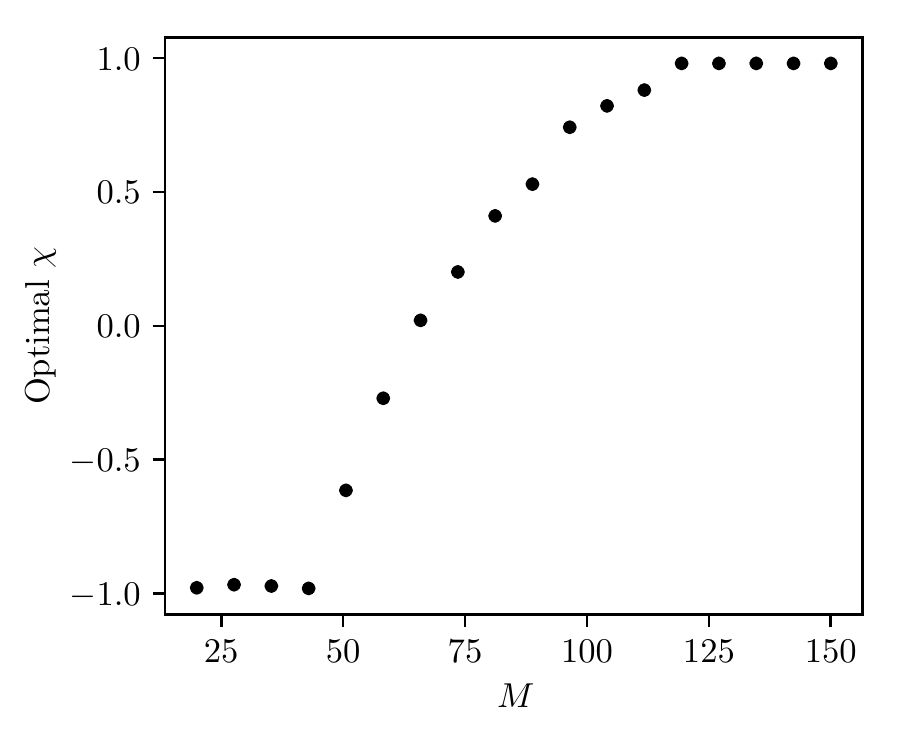}
    \caption{Left panel: the matched correlation of the GW150914 template with waveforms having masses $m_1$ and $m_2$ and spins optimized to extend the degeneracy. The GW150914 masses and uncertainties are indicated. Right panel: the value of $\chi$ at which the cross-correlation is maximized for some sample templates with equal component masses and total mass $M$.}
    \label{fig:7}
\end{figure}

In the preceding subsections, we have identified degeneracy in three parameter cross sections.
Now we consider the more general case, where for each pair of masses $m_1$ and $m_2$, we vary $\chi_1$ and $\chi_2$ to maximize the cross-correlation with the reference waveform, in a 200 ms window.
As we see in the left panel of figure \ref{fig:7}, spin extends the degeneracy in $m_1$ and $m_2$ considerably.
Moreover, for each pair of masses, we record the optimal values of $\chi_1$ and $\chi_2$ and compute the effective spin $\chi$, defined in eq. \ref{eq:chi}.
The optimal $\chi$ increases with the binary mass $M$ as far as possible.
This is illustrated in the right panel of figure \ref{fig:7}, where we choose 18 equal mass waveforms with various total masses $M$ and plot the value of $\chi$ at which the cross-correlation with the GW150914 template is maximized.
Evidently, an increase in $M$ can be compensated by an increase in $\chi$, and vice versa.
This result can be understood because, roughly speaking, an aligned ($\chi > 0$) spin-orbit coupling and an increased $M$ have opposite effects on the frequency evolution; similarly, an anti-aligned ($\chi < 0$) spin-orbit coupling and a decrease in $M$ work in opposite directions \cite{Campanelli}.

\section{Extension of the length of waveforms}
\label{sec:6}

In section \ref{sec:5}, all waveforms were bandpassed from 35--350 Hz and cropped for 100 ms or 200 ms, values representative of the GW150914 event.
Here we perform two tests in order to investigate the effect of the bandpass on the waveform morphology.
Lower frequencies dominate during the earlier parts of the waveform, and in order to capture these effects, here we always use 200 ms, taking 170 ms before the peak and 30 ms after the peak in the ringdown domain.

\begin{figure}[tbp]
    \centerline{
    \includegraphics[scale=0.8]{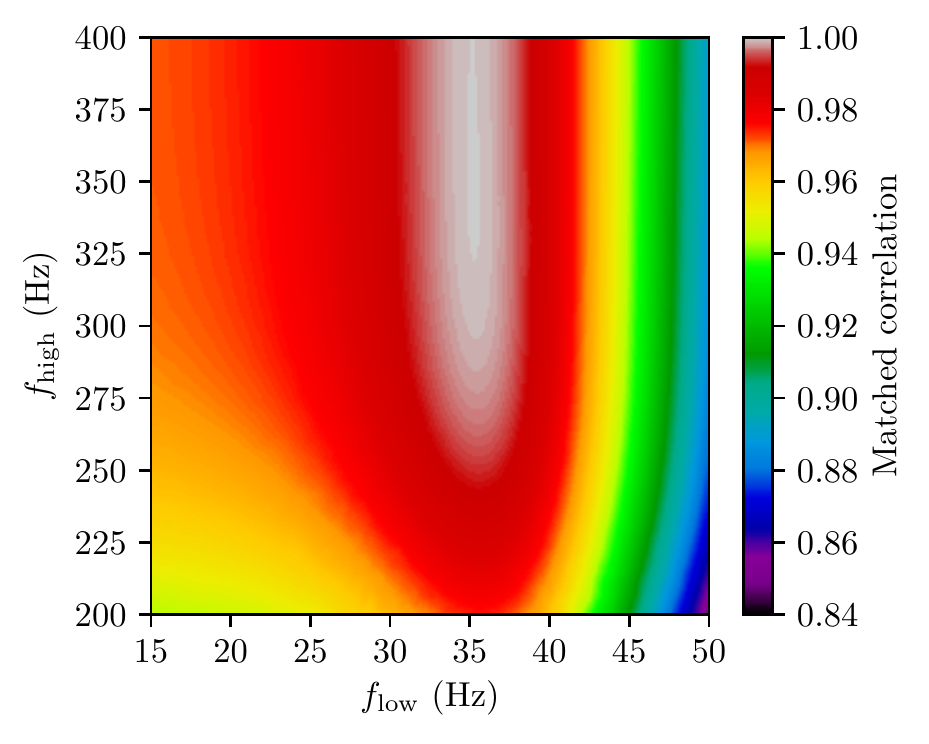}
    \includegraphics[scale=0.8]{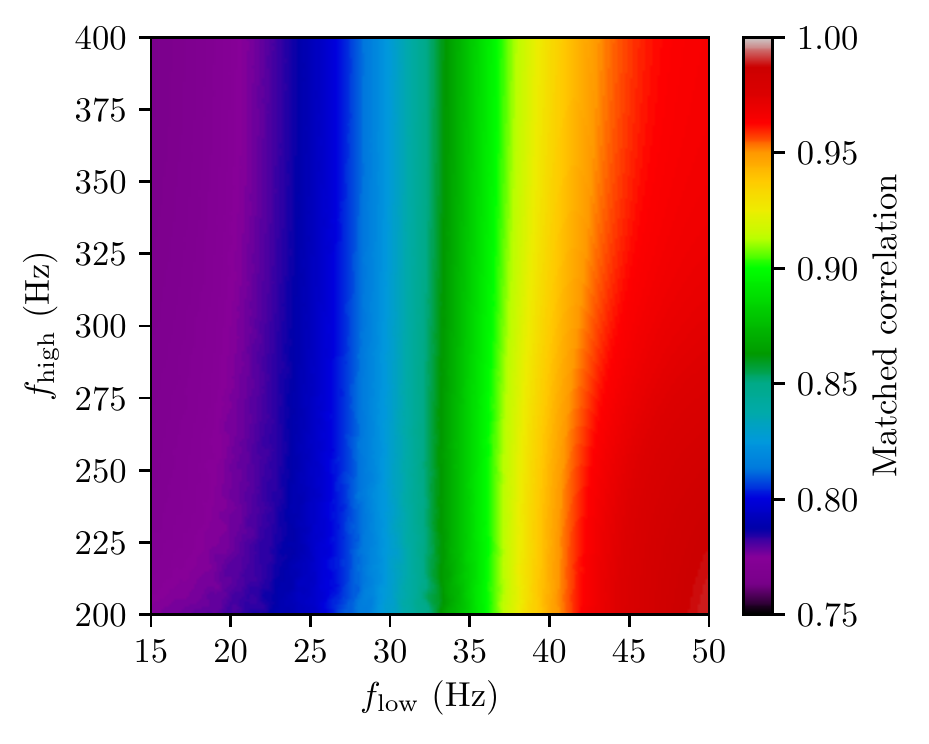} } 
    \caption{Left panel: the matched correlation of the GW150914 waveform, bandpassed from 35--350 Hz, with the same waveform bandpassed from $f_{\mathrm{low}}$--$f_{\mathrm{high}}$. Right panel: the matched correlation of the original GW150914 waveform with the a new waveform with $m_1 = 50, m_2 = 35, \chi_1 = 0.95, \chi_2 = -0.95$, as a function of the bandpass frequencies used on both waveforms.} 
    \label{fig:8}
\end{figure}

First, we take the reference template with $m_1 = 36$, $m_2 = 29$, $\chi_1 = \chi_2 = 0$, apply a bandpass filter with different pass bands, and compute the matched correlation with the same template bandpassed from 35--350 Hz.
The results are shown in the left panel of figure~\ref{fig:8}.
The matched correlation is insensitive to $f_\mathrm{high}$, and it falls off slowly with decreasing $f_\mathrm{low}$ and falls off somewhat faster with increasing $f_\mathrm{low}$.

In the second test, we take the original GW150914 template with $m_1 = 36$, $m_2 = 29$, $\chi_1 = \chi_2 = 0$ and a new template with $m_1 = 50$, $m_2 = 35$, $\chi_1 = 0.95, \chi_2 = -0.95$, both filtered by the same bandpass filter with various low and high frequencies, and compute their matched correlation.
The results are shown in the right panel of figure \ref{fig:8}.
Again, the matched correlation depends weakly on $f_\mathrm{high}$, and the correlation slightly improves with increasing $f_\mathrm{low}$.
Decreasing $f_\mathrm{low}$ from 35 Hz down to lower frequencies reduces the degeneracy considerably, with the cross-correlation falling from nearly 0.9 to below 0.8 at 15 Hz.

These tests suggest that frequencies outside of 35--350 Hz do not drastically affect the morphology for masses in the vicinity of GW150914 and within the 200 ms region under consideration. 
Within the 100 ms region, the effect of the bandpass on morphology is even weaker.

\section{Signal-to-noise ratio and waveform degeneracy}
\label{sec:7}

A peak in the signal-to-noise ratio (SNR) identifies the presence of a certain template within noisy data. 
SNR peaks are used by LIGO to detect gravitational waves \cite{Allen}.
Because of the degeneracy of gravitational waveforms, the SNR peak for GW150914 is not very sensitive to the masses and spins of the proposed template.
To illustrate this we consider two arbitrary templates with masses drastically different from the median masses of 36 and 29:
\begin{itemize}
    \item $m_1 = 70$, $m_2 = 35$, $\chi_1 = \chi_2 = 0$
    \item $m_1 = 70$, $m_2 = 35$, $\chi_1 = -\chi_2 = 0.95$
\end{itemize}

We use each of these two templates to calculate the SNR within the Hanford and Livingston 32 second cleaned data files that are centred on the GW150914 event.
The resulting peak is compared with the peak given by the spinless GW150914 template, with $m_1 = 36, m_2 =29$, and $\chi_1 = \chi_2 = 0$.

\begin{figure}[tbp]
    \centering
    \includegraphics[scale=0.48]{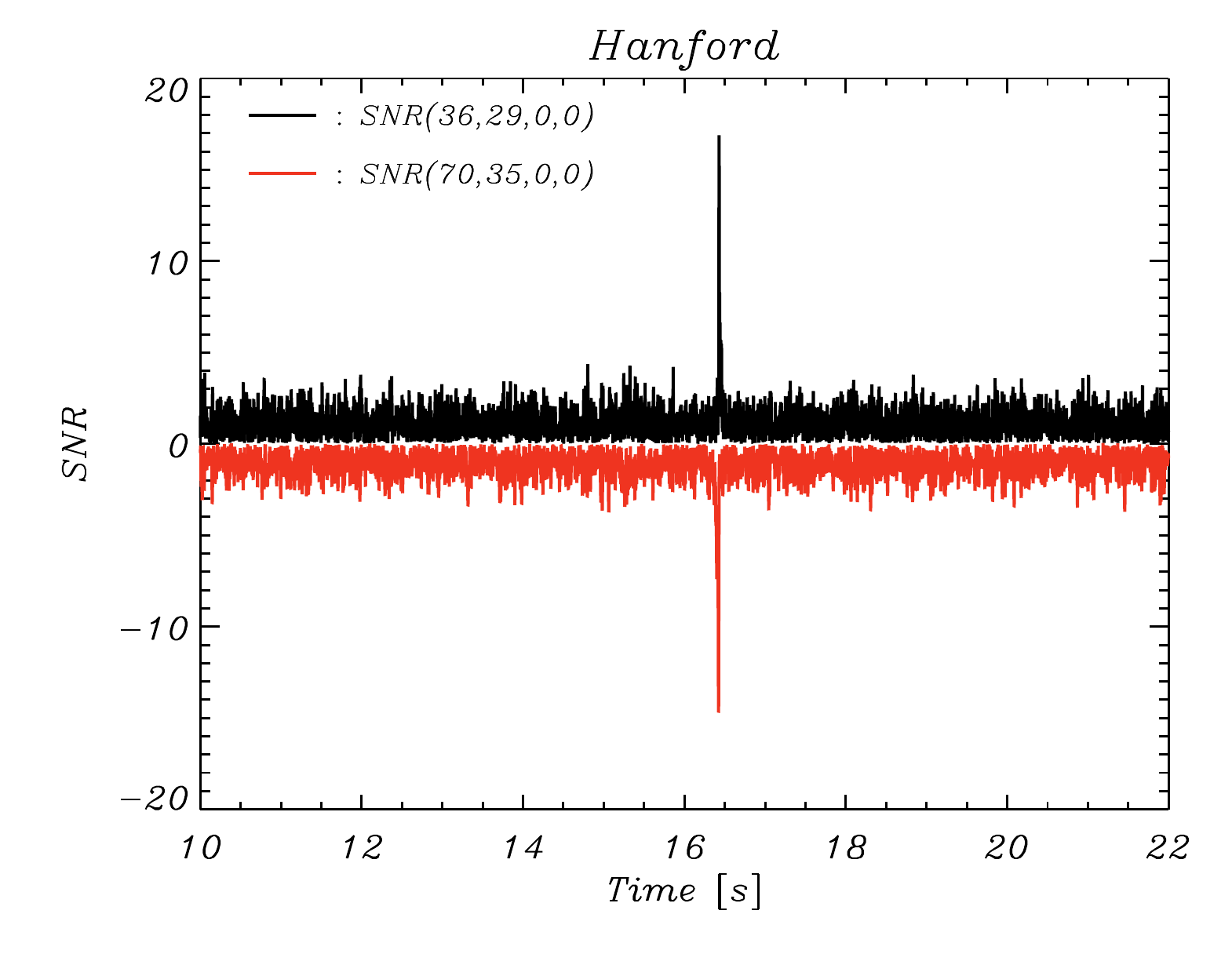}
    \includegraphics[scale=0.48]{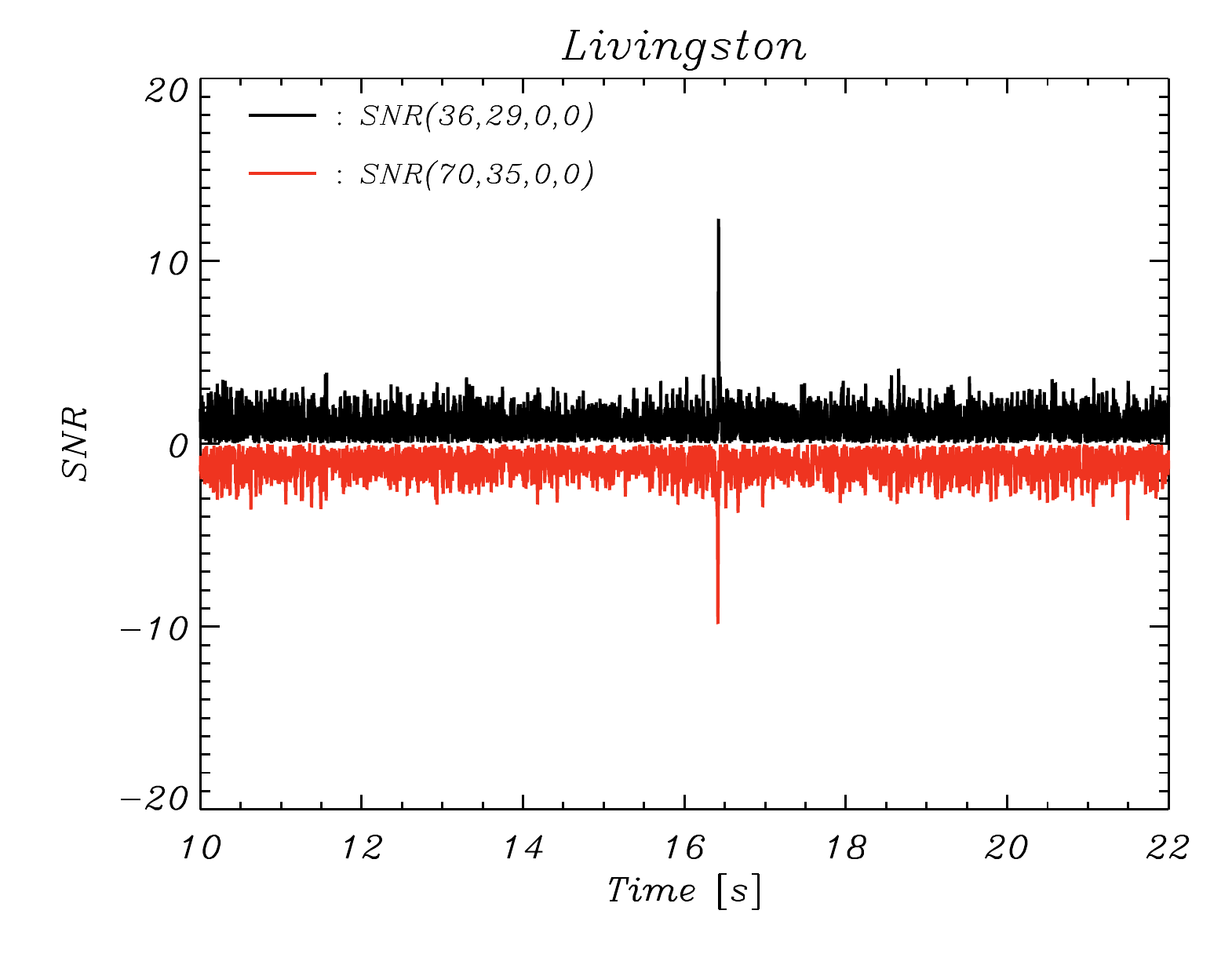}
    \includegraphics[scale=0.48]{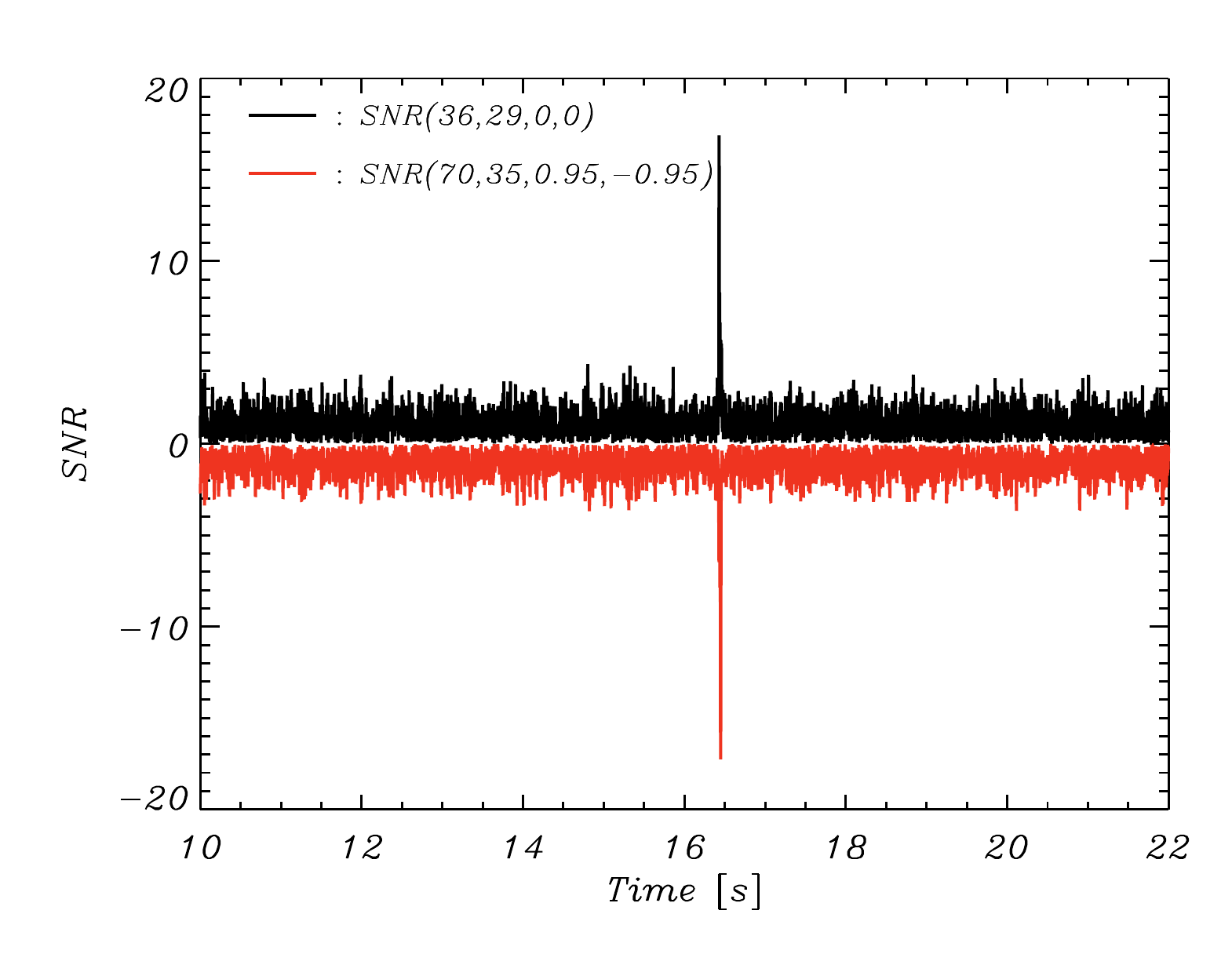}
    \includegraphics[scale=0.48]{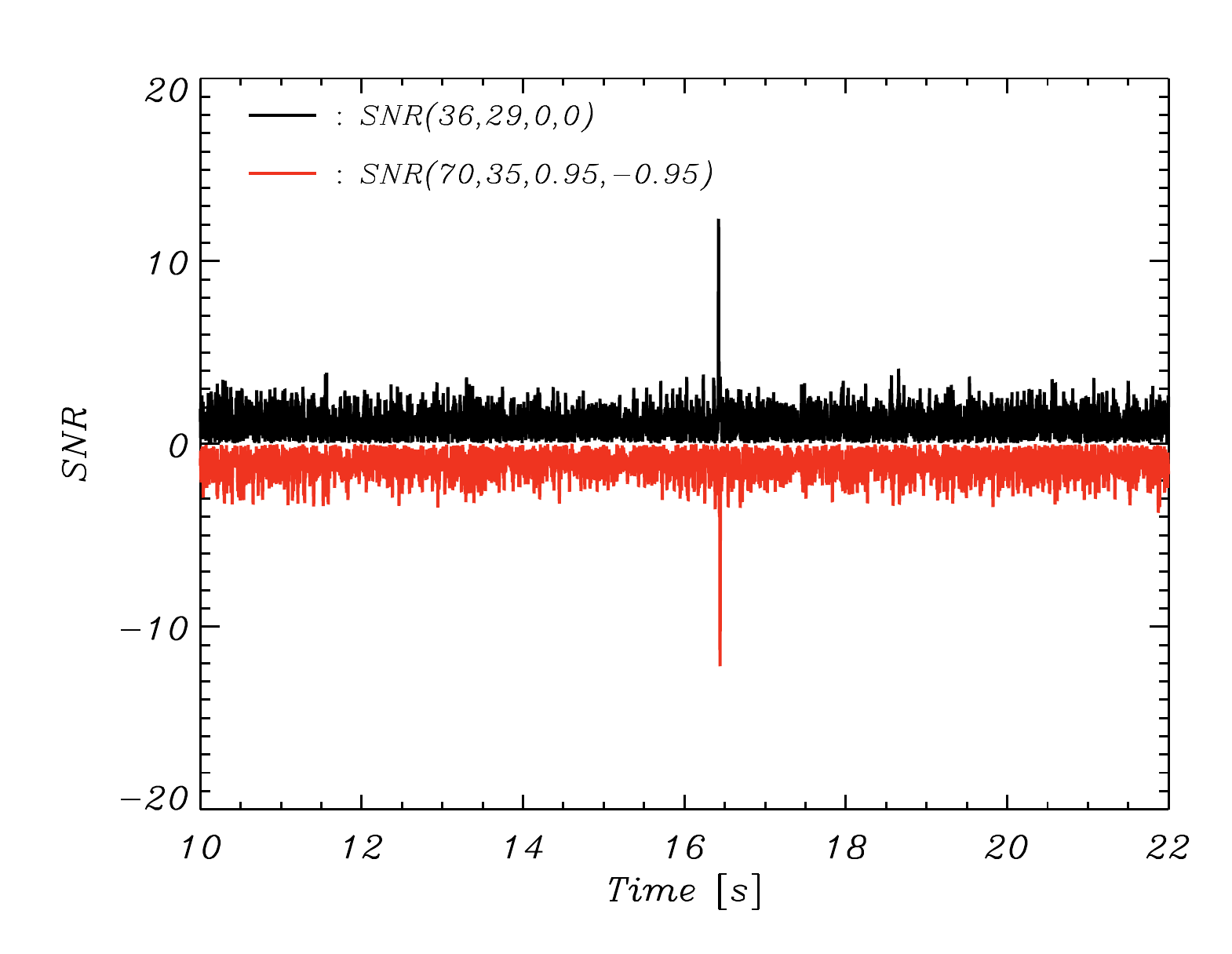}
    \caption{The signal-to-noise ratio $\mathrm{SNR}(m_1, m_2, \chi_1, \chi_2)$ plotted for the reference template (in black) and the alternate templates (inverted, in red) with the data from Hanford (left) and Livingston (right). Here we used \texttt{PyCBC} with whitening, but no bandpassing.}
    \label{fig:9}
\end{figure}

As the results in figure \ref{fig:9} show, both alternate templates give a significant SNR peak at GW150914.
The SNR detection method does not distinguish between templates with masses differing by a factor of 2.
In the case of the $70, 35, 0.95, -0.95$ template, the peak is actually slightly stronger than the reference template. 
This template is plotted in figure \ref{fig:11} after matching to the spinless GW150914 reference template.
Here we also plot the template with masses $m_1=48, m_2=37$ and strong anti-aligned spins, which maximizes the SNR.

\begin{figure}[tbp]
    \centerline{
    \includegraphics[scale=0.5]{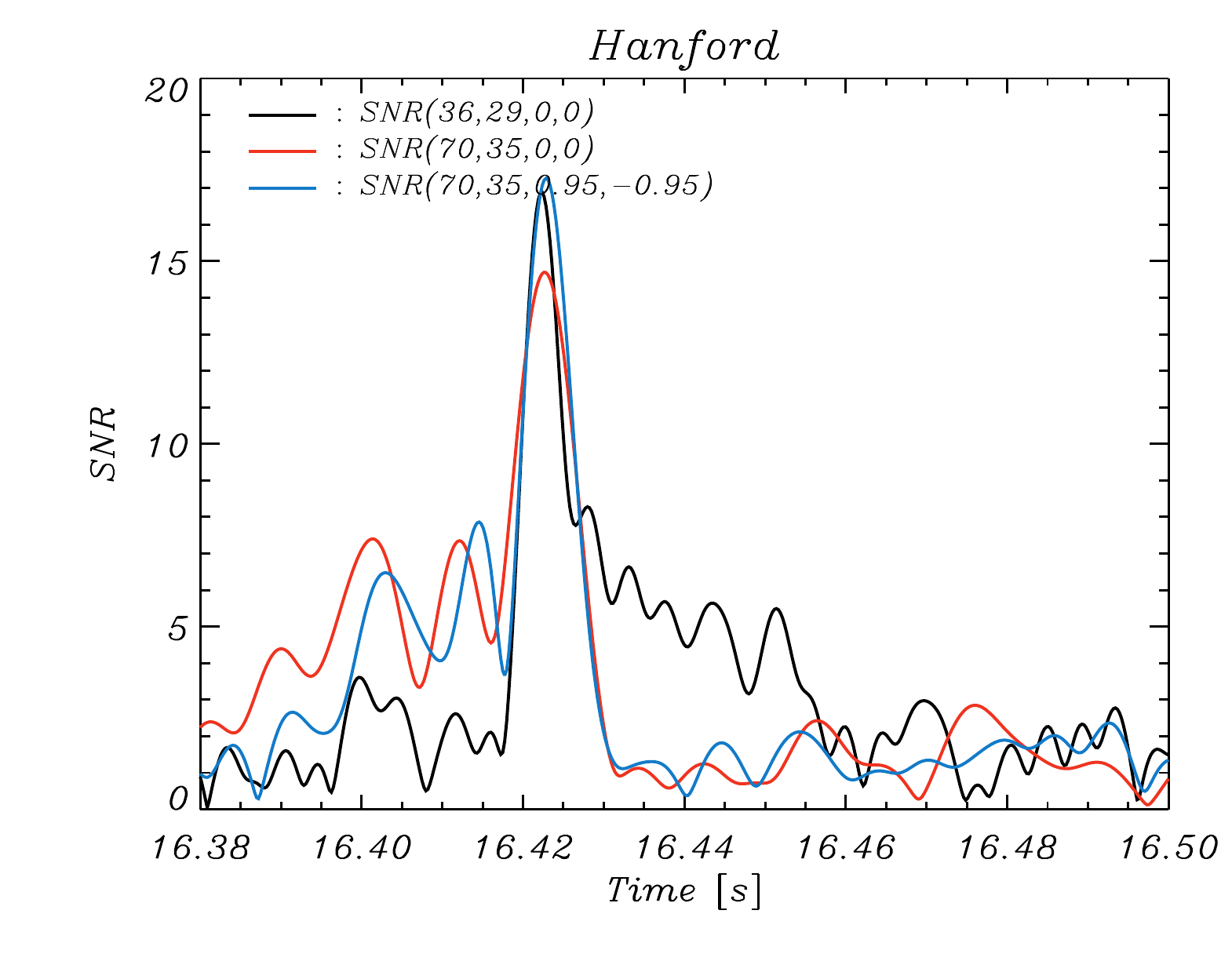}
     \includegraphics[scale=0.5]{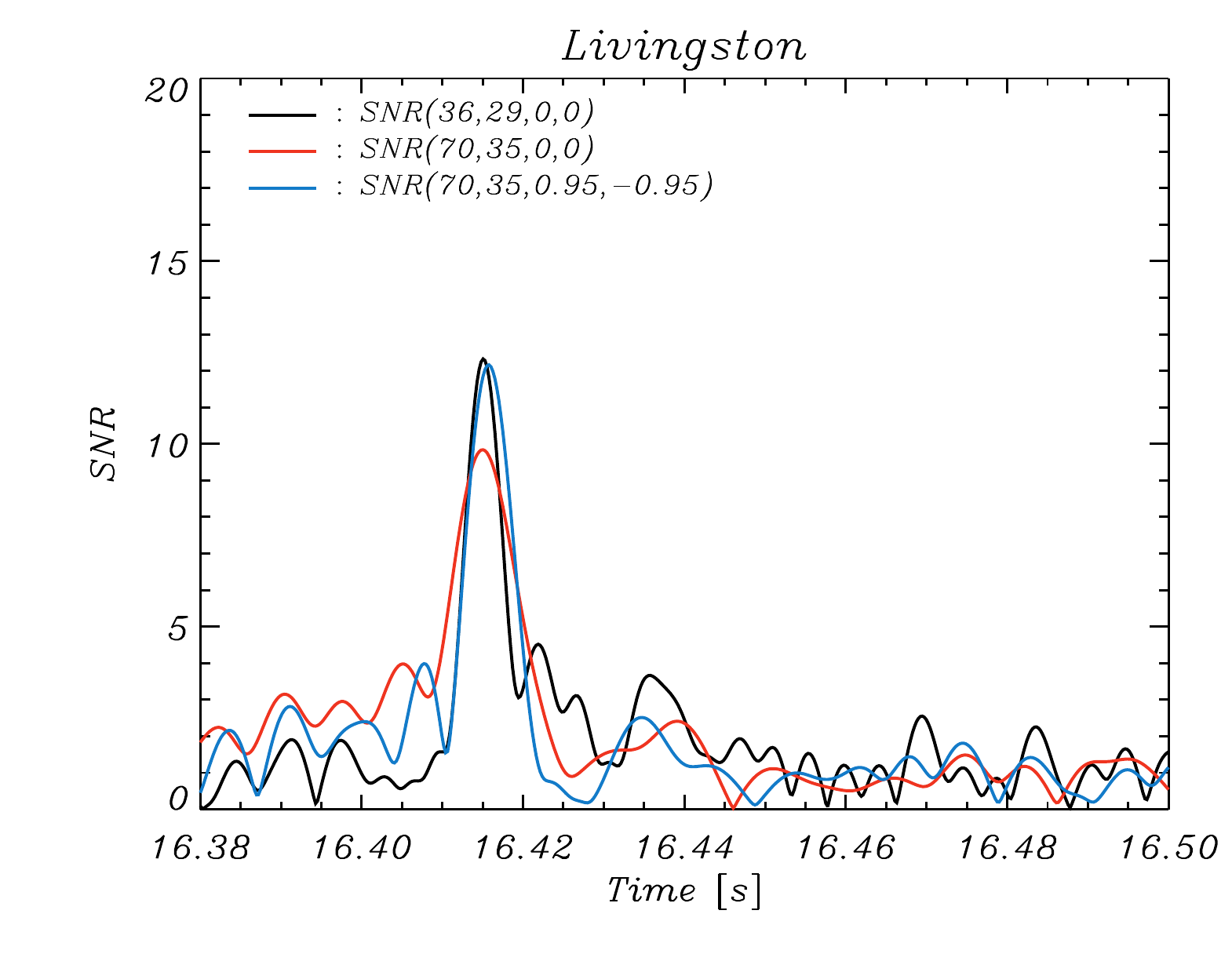}} 
    \caption{Zoomed-in signal-to-noise ratio $\mathrm{SNR}(m_1, m_2, \chi_1, \chi_2)$ plotted in the vicinity of the points of global maximum for the reference template (in black) and the alternate templates (in red) with the data from Hanford (left) and Livingston (right).}
    \label{fig:10}
\end{figure}

\begin{figure}[tbp]
    \centering
    \includegraphics[width=0.48\textwidth]{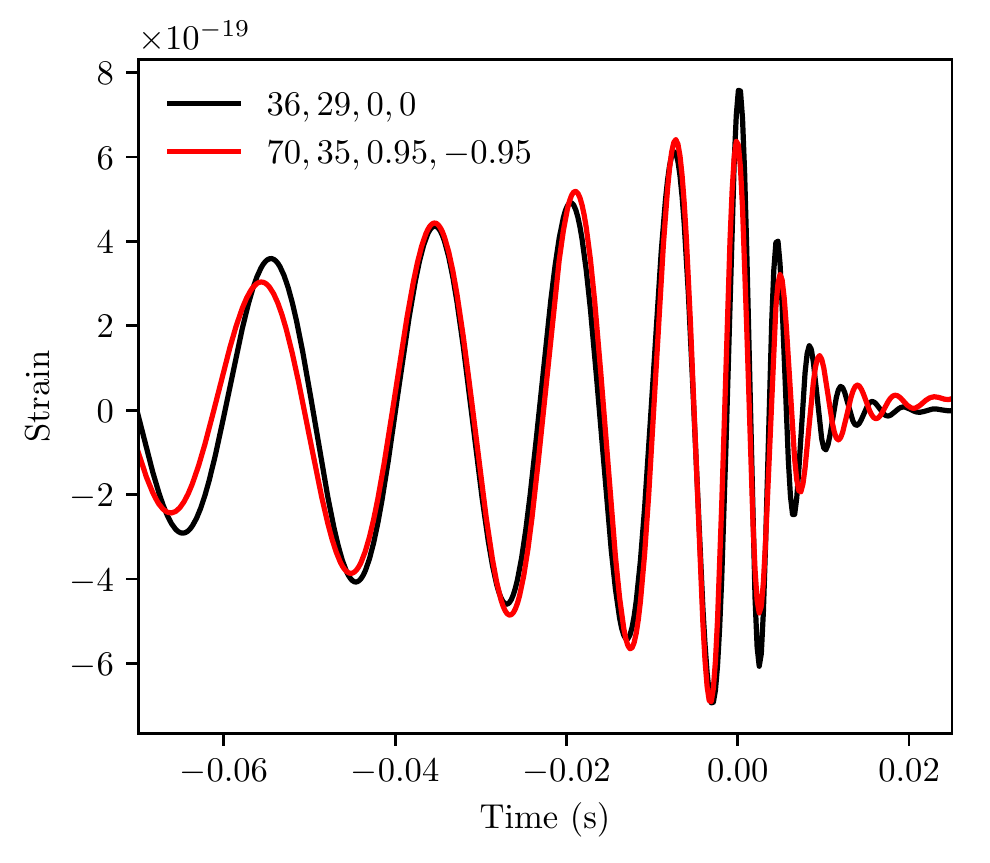}
    \includegraphics[width=0.48\textwidth]{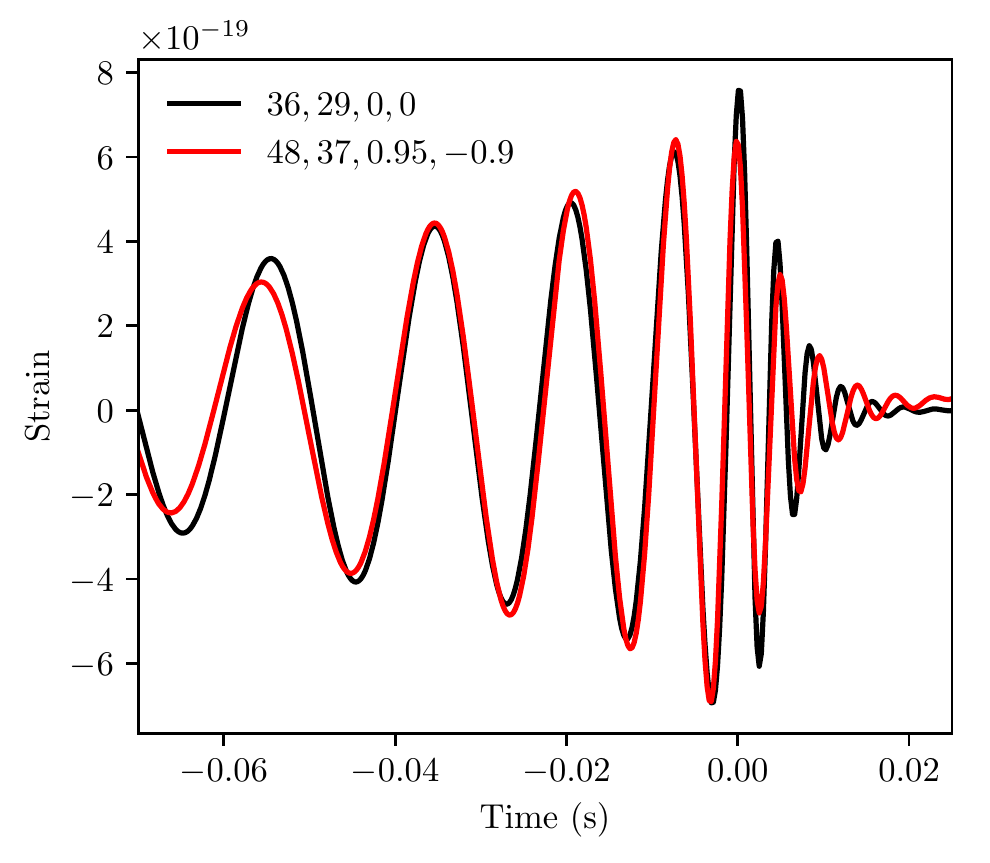}
    \caption{Left panel: the $m_1 = 70$, $m_2 = 35$, $\chi_1 = 0.95$, $\chi_2 = -0.95$ template matched to the reference template. This template lies slightly outside the region of strongest degeneracy, however it is still similar in morphology (cross-correlation is $0.91$) and it triggers an equally significant SNR peak in the Hanford and Livingston data (see figure \ref{fig:10}). Right panel: the maximum SNR template with $m_1 = 48$, $m_2 = 37$, $\chi_1 = 0.95$, $\chi_2 = -0.9$ is extremely similar to the reference template (cross-correlation exceeds $0.96$), although its masses lie outside the credible regions.}
    \label{fig:11}
\end{figure}

\section{Limitations on parameter estimation in the presence of noise}
\label{sec:8}

The noise-free analysis in section \ref{sec:5} reveals the shape of the degeneracy in the vicinity of parameters similar to those of GW150914. 
In the practical problem of parameter extraction, one is concerned with the accuracy with which waveforms can be resolved and parameters determined in the presence of detector noise.

In this section, we estimate the constraints on parameter accuracy in the context of the method developed in ref.~\cite{Liu2}.
This template-free method, which makes no assumptions about noise properties, constructs the common signal in two or more detectors by maximizing the cross-correlation with the strain data and minimizing the cross-correlation between the residuals.
The method does not identify a unique best-fit template, but rather a family of best-fit templates, differing from each other, and from theoretical gravitational wave templates, because of chance correlations in the strain data. 
The degree to which different members of the family differ from each other---and the corresponding ambiguity in the extracted signal---reflects the inherent uncertainty in the problem due to the presence of noise.
As the signal becomes stronger or the noise weaker, the family of best-fit templates converges to the true signal.

The typical cross-correlation between members of the family of blindly-extracted templates sets an approximate threshold, above which templates cannot be reliably distinguished and are effectively equivalent, and below which templates can be distinguished and rejected.
To determine this threshold for GW150914, we perform the blind search method and extract 100 templates, using the same time (200 ms) and frequencies (35--350 Hz) as in the present paper.
The resulting distribution of cross-correlations between pairs of templates is tightly centred around 0.90, with none below 0.88.
Therefore, in the cross-correlation plots in figures \ref{fig:4}--\ref{fig:7}, the contours corresponding to $\approx0.88$ enclose the approximate ``credible region'', outside which we can exclude parameters as correct for GW150914, but within which no further distinction can be made.
With the colour map in use in all plots, this is approximately the red region.
For example, in the Schwarzschild case (figure~\ref{fig:4}), the allowed region crosses the entire 10--75$M_\odot$ range under consideration.

The uncertainties derived in our approach are more conservative than those from Bayesian inference as a consequence of the fact that we are unwilling to introduce assumptions regarding the nature of the noise (i.e., stationarity and Gaussianity) that are known to be incorrect.
Furthermore, we constrain ourselves to use data exclusively from the time and frequency domains where the signal is visible.
LIGO acknowledges that the noise is neither Gaussian nor stationary\footnote{``The usual interpretation of our credible intervals relies on the assumption that both
our signal and noise model are an appropriate description of the data. The previous
section addressed the signal model, but the zero-noise method does not take into account
the properties of actual detector noise, such as non-Gaussianity, non-stationarity and
inaccuracies in PSD estimations.'' \cite{Ligo5}}. Since stationarity and Gaussianity are prerequisites for the Bayesian analysis, LIGO's Bayesian methods---including eqs.~(\ref{eq:likelihood}) and (\ref{eq:rho}) above---are without theoretical support and their results cannot be regarded as reliable.

We also note that the argument leading to a threshold of distinguishability of 0.88 is qualitatively reliable but not necessarily quantitatively correct. The choice of a reference template with $m_1 = 36$ and $m_2 = 29$ was prompted by the values given in ref.~\cite{Ligo2}.
A similar degree of degeneracy is to be expected for other BBH parameters.
Nevertheless, it is clear that there is strong disagreement between the results of LIGO's Bayesian inference and the present assumption-free method. 
Specifically, our results suggest that LIGO's mass uncertainties are too small by a factor of approximately 5.

\section{Discussion and conclusion}
\label{sec:9}

Accurate extraction of the masses and spins from a BBH waveform has important applications in astrophysics and cosmology, including understanding black hole formation. 
However, the degeneracy of gravitational waveforms can impose limitations on the accuracy with which parameters can be extracted, especially when only a short duration of usable signal is present.
In order to investigate these limitations in the context of GW150914, we have studied the degeneracy surrounding the GW150914 masses $m_1 = 36$, $m_2 = 29$.

To leading order there is the chirp mass degeneracy between $m_1$ and $m_2$.
The inclusion of higher order terms, as well as the cleaning procedure, including bandpassing, cropping, and matched filtering, widens this line into a basin of induced degeneracy spreading across mass space.

We have also allowed the black holes to have spins aligned with the orbit axis.
These spins are poorly constrained by the waveform morphology.
Furthermore, binaries with positively aligned spin mimic lower mass non-spinning binaries, and similarly for negatively aligned spin and higher mass.

After cleaning and matching, the cross-correlation between Hanford and Livingston strain data at GW150914 is 0.77 in a 200ms window and 0.87 in a 100ms window \cite{Naselsky}.
It is reasonable to say that our techniques are unable to rule out templates whose cross-correlation with the ``best fit'' template exceeds these values.
The corresponding contours, for example the entire red region in figure \ref{fig:5}, enclose a large region of degeneracy in parameter space extending far beyond the statistical uncertainties obtained in LIGO's parameter estimation \cite{Ligo2}.

It is clear that the arbitrary time shifts and phase shifts introduced here make important contributions to the induced degeneracy.  The parameters appearing in the matched filter are all physical.  The time lag is determined by the sky location, and the constant phase shift is determined by the orientation of the plane of the binary system.  As we have shown, the price of this freedom is degeneracy.  
In the case of GW150914, this means that models of the BBH system with, for example, $m_1 = 48$, $m_2 = 37$, $\chi_1 = 0.95$, $\chi_2 = -0.9$ or with $m_1 = 70$, $m_2 = 35$, $\chi_1 = 0.95$, $\chi_2 = -0.95$ are as good as the template given in ref.~\cite{Ligo1}.  Evidently, it is worth considering the extent to which this degeneracy can be reduced by the independent measurements of these BBH parameters.  We note, however, that a potential increase in the BBH masses would raise fundamental questions regarding their origin.  

\acknowledgments

This work has made use of LIGO software and data, including that obtained from the LIGO Open Science Center (\texttt{https://losc.ligo.org}), a service of LIGO Laboratory, the LIGO Scientific Collaboration and the Virgo Collaboration.
LIGO is funded by the U.S. National Science Foundation.
Virgo is funded by the French Centre National de Recherche Scientifique (CNRS), the Italian Istituto Nazionale della Fisica Nucleare (INFN) and the Dutch Nikhef, with contributions by Polish and Hungarian institutes.

Our research was funded in part by the Danish National Research Foundation (DNRF) and by Villum Fonden through the Deep Space project.
Hao Liu is supported by the Youth Innovation Promotion Association, CAS.

\end{document}